\documentclass[aps,prl,twocolumn,amsmath,showpacs]{revtex4}%superscriptaddress
\usepackage{bm}
\usepackage{graphicx}
\begin{document}
\setlength{\arraycolsep}{2pt}

%\begin{document}

\title{Loophole-free Bell test for continuous variables via wave and particle correlations}
\author{Se-Wan Ji$^{1,2}$, Jaewan Kim$^{2}$, Hai-Woong Lee$^{3}$, M. S. Zubairy$^{4}$, and Hyunchul Nha$^{1,2,*}$} %$^*$
\affiliation{$^1$Department of Physics, Texas A \& M University at Qatar, Doha, Qatar\\
$^2$School of Computational Sciences, Korea Institute for Advanced Study, Seoul 130-012, Korea\\
$^3$Department of Physics, Korea Advanced Institute of Science and
Technology, Daejeon 305-701, Korea\\
$^4$Department of Physics and Institute of Quantum Studies, Texas A\& M University, College Station, TX 77843, USA}

%\author{Se-Wan Ji}
%\affiliation{Department of Physics, Texas A $\&$ M University at Qatar, PO Box 23874, Doha, Qatar}
%\affiliation{School of Computational Sciences, Korea Institute for Advanced Study, Seoul 130-012, Korea}
%\author{Jaewan Kim}
%\affiliation{School of Computational Sciences, Korea Institute for Advanced Study, Seoul 130-012, Korea}
%\author{Hai-Woong Lee}
%\affiliation{Department of Physics, Korea Advanced Institute of Science and
%Technology, Daejeon 305-701, Korea}
%\author{M. S. Zubairy}
%\affiliation{Department of Physics and Institute of Quantum Studies, Texas A $\&$ M University, College Station, TX 77843, USA}
%\author{Hyunchul Nha}$^*$
%\email{hyunchul.nha@qatar.tamu.edu }
%\affiliation{Department of Physics, Texas A $\&$ M University at Qatar, PO Box 23874, Doha, Qatar}
%\affiliation{School of Computational Sciences, Korea Institute for Advanced Study, Seoul 130-012, Korea}
\date{\today}

\begin{abstract}
We derive two classes of multi-mode Bell inequalities under local realistic assumptions, which are violated only by the entangled states negative under partial transposition in accordance with the Peres conjecture. 
Remarkably, the failure of local realism can be manifested by exploiting wave and particle correlations of readily accessible continuous-variable states, with very large violation of inequalities insensitive to detector-efficiency, which makes a strong case for a loophole-free test. 
\end{abstract}

\pacs{03.65.Ud, 03.65.Ta, 03.67.Mn, 42.50.Dv}
\maketitle

{\it Introduction}---Whether there exists a strong correlation that no local realistic theories can grasp has been an issue of crucial importance 
since the Einstein-Podolsky-Rosen (EPR) argument \cite{EPR}, which was later cast into an experimentally testable form by J. S. Bell \cite{Bell}.  
The Bell test not only provides an opportunity to look into fundamental aspects of quantum mechanics, but also can be used for practical applications in quantum information science, e.g., the security test for quantum cryptography \cite{Ekert} and the entanglement witness.
Numerous experimental data have been obtained to date in support of quantum mechanics, however, there still remain some important issues to resolve. 
First, no experiment ever closed both the locality and the detector-efficiency loopholes to conclusively rule out local hidden variable (LHV) theories \cite{Weihs,Fry}. 
Second, although the original EPR argument considered the correlation of continuous variables (position and momentum), 
almost all experiments were so far performed for discrete variables (e.g. spin-1/2 states \cite{Ansmann}). 

The Bell test using continuous variables (CVs) can provide a new insight into quantum world via their enriched structure in infinite dimension. 
Furthermore, the CV Bell test is considered practically desirable for a loophole-free test because the measurement scheme (homodyne detection) is highly efficient. 
However, the proposals so far have not been made to take the merits of CVs fully.
To begin with, the EPR state (two-mode squeezed state) is not adequate as such for the CV Bell test due to a non-negative distribution in phase space, admitting a LHV description \cite{Bell1}. It was thus suggested to exploit the correlation of discrete nature, photon-number parity \cite{Banaszek} or pseudo-spin observables \cite{Chen}, which are hard to implement due to inefficient photon counting. 
In order to utilize the merit of homodyne detection, a different approach, i.e. transforming a nonnegative distribution to a nonpositive one by photon subtraction, was proposed  but the violation of Bell inequality was very small \cite{Nha1}. 
This small violation may be attributed to the binning process that converts CV data to binary ones; 
binning is used to adopt the Bell-inequalities typically established for discrete variables \cite{Bell}, leading undesirably to the loss of information on CV correlation.  
Remarkably, Acin {\it et al.} found some CV states that maximally violate multi-mode Bell inequalities under the binning and 
proposed a 3-mode state for a loophole-free test, which however is not very practicable in current technologies \cite{Acin0}.

Therefore, it is important both fundamentally and practically to have Bell-inequalities that can directly probe CV correlations in full capacity \cite{Vogel}. 
Recently, such an inequality was derived by Cavalcanti {\it et al.} \cite{Cavalcanti}, however, its test appears demanding as it requires at least 10-mode entangled states.
Although the case was improved to use 5-mode states by optimizing the functional form of the inequalities \cite{He}, 
it is necessary to obtain Bell inequalities that can reveal nonlocality for a broad class of CV entangled states including, desirably, easily accessible ones.
In this Letter, we derive two classes of Bell inequalities by using the Cauchy inequality under local realistic conditions. 
We show that these inequalities can be violated only by the quantum entangled states that are negative under partial transposition (NPT), in accordance with the Peres conjecture \cite{Peres,Acin, Nha2}. Remarkably, the violation of our inequalities occurs at all levels of  $n$-mode ($n\ge2$), 
illustrated by well-known, readily accessible, two-mode entangled states. % (single-photon entangled state, two-mode EPR state, and entangled coherent state). 
Our inequalities require the comparison of two distinct correlations, wave-like (homodyne detection) and particle-like (photon-counting) correlations. 
We show that the degree of violation can be very large, and furthermore, that our tests are insensitive to detector-efficiency, therefore suitable for a loophole-free test within existing technologies. 

%Before starting, let us first note that the problem of quantum nonlocality must be clearly distinguished from that of entanglement detection. 
%The latter issue is to ask {\it within quantum formalism} whether a given state can be represented by a convex sum of product states, $\rho=\sum_ip_i\rho_i^A\otimes\rho_i^B$ (separable state). On the other hand, the former issue is to distinguish between quantum and classical formalisms. 
%Thus, it is important to derive a Bell inequality entirely in terms of classical principles. 
%In this respect, there are some previous works that were not based on a valid classical method with no rigorous context of Bell theorem.

{\it Bell inequality}---We first show how a Bell inequality can be derived from LHV descriptions. 
Let $r_j$ be a real random variable  at two parties $j=1,2$. 
The LHV theory accounts for the correlation of $r_1$ and $r_2$ by 
\begin{eqnarray}
\langle r_1r_2\rangle=\int d\lambda\rho(\lambda)r_1(\lambda)r_2(\lambda),
\label{eqn:LHVR}
\end{eqnarray}
where it is assumed that the local values $r_1(\lambda)$ and $r_2(\lambda)$ are predetermined (realism) independent of each party (locality). 
The realistic values $r_1(\lambda)$ and $r_2(\lambda)$ can be identified if the hidden variable $\lambda$, with the probabilistic distribution $\rho(\lambda$), is revealed.
This can be extended to complex variables $C_1$ and $C_2$ that essentially represent two real random variables at each site,  as
\begin{eqnarray}
\langle C_1C_2\rangle=\int d\lambda\rho(\lambda)C_1(\lambda)C_2(\lambda),
\label{eqn:LHVC}
\end{eqnarray} 
which refer to four correlations collectively [Cf. Eq. (9)].
We impose no conditions on random variables that may be bounded/unbounded and continuous/discrete. 

One can use the Cauchy inequality to obtain the upper bound of the correlation as
\begin{eqnarray}
 %\left|\int d\lambda\rho(\lambda)C_1(\lambda)C_2(\lambda)\right|^2\nonumber\\
&&|\langle C_1C_2\rangle|^2=\left|\int d\lambda\rho(\lambda)C_1^p(\lambda)C_2^q(\lambda)C_1^{1-p}(\lambda)C_2^{1-q}(\lambda)\right|^2\nonumber\\
&&\le\int d\lambda\rho(\lambda)\left|C_1^p(\lambda)C_2^q(\lambda)\right|^2\int d\lambda\rho(\lambda)\left|C_1^{1-p}(\lambda)C_2^{1-q}(\lambda)\right|^2\nonumber\\
&&=\langle \left|C_1\right|^{2p}\left|C_2\right|^{2q}\rangle\langle \left|C_1\right|^{2(1-p)}\left|C_2\right|^{2(1-q)}\rangle, \nonumber
\end{eqnarray}
where the last line again follows in view of  the LHV description in Eq.~(\ref{eqn:LHVR}), with real numbers $p, q\in[0,1]$.
Therefore, we obtain a Bell inequality
\begin{eqnarray}
\left|\langle C_1C_2\rangle\right|^2\le \langle \left|C_1\right|^{2p}\left|C_2\right|^{2q}\rangle\langle \left|C_1\right|^{2(1-p)}\left|C_2\right|^{2(1-q)}\rangle.
\label{eqn:Bell1}
\end{eqnarray}
If one starts with the complex conjugate $C_2^*$ instead of $C_2$, another inequality similarly emerges,
\begin{eqnarray}
\left|\langle C_1C_2^*\rangle\right|^2\le\langle \left|C_1\right|^{2p}\left|C_2\right|^{2q}\rangle\langle \left|C_1\right|^{2(1-p)}\left|C_2\right|^{2(1-q)}\rangle.
\label{eqn:Bell2}
\end{eqnarray}

In~(\ref{eqn:Bell1}) and (\ref{eqn:Bell2}), 
the correlation of $C_1$ and $C_2$ is bounded from above. %the complex variables 
We particularly note that considering $C_1$ and $C_2$ as complex amplitudes, the upper bound---the product of $\langle\left|C_1\right|^{2p}\left|C_2\right|^{2q}\rangle$ and $\langle\left|C_1\right|^{2(1-p)}\left|C_2\right|^{2(1-q)}\rangle$---
refers to the (fractional-order) ``intensity" correlations. 

Now, we want to know if the inequalities ~(\ref{eqn:Bell1}--\ref{eqn:Bell2}) can be violated by quantum systems. 
We first discuss how the correlations in our inequalities can be experimentally tested. 
The left-hand side (LHS) of each inequality refers to the correlation of complex amplitudes $C_j\equiv C_{jx}+iC_{jy}$ ($j=1,2$), 
which can be addressed in quantum mechanics by introducing the operators $\hat{C}_j\equiv\hat{C}_{jx}+i\hat{C}_{jy}$ 
($\hat{C}_{jx},\hat{C}_{jy}$: Hermitian).
On the other hand, the right-hand side (RHS) of each inequality refers to the intensity correlation.
An intensity can generally be expressed in two different forms, which, importantly, are not distinguished from each other in classical descriptions. 
First is to represent a complex variable by its real and imaginary parts, $C\equiv C_x+iC_y$,  leading to $|C|^2=C_x^2+C_y^2$. 
Second is to represent the intensity as the product of the original variable and its conjugate, $|C|^2=C^*C$.  
The former indicates the correspondence to quantum operator as $\left|C\right|^2\doteq \hat{C}_x^2+\hat{C}_y^2$ and the latter 
$\left|C\right|^2\doteq \hat{C}^\dag\hat{C}$. 

Two unequal observables ( $\hat{C}_x^2+\hat{C}_y^2$ and $\hat{C}^\dag\hat{C}$) in quantum domain usually carry distinguished physical contexts.
The distinction may be particularly related to the wave-particle duality in quantum optics, as addressed below. 
The classical LHV descriptions, however, disallow the violation of inequalities~(\ref{eqn:Bell1}--\ref{eqn:Bell2}) 
regardless of intensity observables. 
Here we particularly focus on the second approach, $\left|C\right|^2\doteq \hat{C}^\dag\hat{C}$. 
For simplicity, let $\{p,q\}=\{0,1\}$, and then, we have only two distinct cases. 

(i) $p=q=1$: 
One class of inequalities follows from~(\ref{eqn:Bell2}),
\begin{eqnarray}
\left|\langle \hat{C}_1\hat{C}_2^\dag\rangle\right|^2\le\langle \hat{C}_1^\dag\hat{C}_1\hat{C}_2^\dag\hat{C}_2\rangle\hspace{0.3cm}:{\rm 1st-inequality}.
\label{eqn:1st-Bell-Q}
\end{eqnarray}
There is another inequality from~(\ref{eqn:Bell1}), $\left|\langle \hat{C}_1\hat{C}_2\rangle\right|^2\le\langle \hat{C}_1^\dag\hat{C}_1\hat{C}_2^\dag\hat{C}_2\rangle$, 
which is, however, never violated by any quantum states as shown below.

(ii) $p=0,q=1$ : Another class follows from~(\ref{eqn:Bell1})
\begin{eqnarray}
\left|\langle \hat{C}_1\hat{C}_2\rangle\right|^2\le\langle \hat{C}_1^\dag\hat{C}_1\rangle\langle \hat{C}_2^\dag\hat{C}_2\rangle\hspace{0.3cm}:{\rm 2nd-inequality}.
\label{eqn:2nd-Bell-Q}
\end{eqnarray}
The other  inequality from~(\ref{eqn:Bell2}), 
$\left|\langle \hat{C}_1\hat{C}_2^\dag\rangle\right|^2\le\langle \hat{C}_1^\dag\hat{C}_1\rangle\langle \hat{C}_2^\dag\hat{C}_2\rangle$, is never violated as shown below. 

{\it Peres Conjecture}---We now prove that only NPT entangled states can violate the Bell inequalities~(\ref{eqn:1st-Bell-Q}) and~(\ref{eqn:2nd-Bell-Q}) regardless of  $\hat{C}_1$ and $\hat{C}_2$. First, note that for any operator $\hat{f}$, the positive operator $\hat{f}^\dag\hat{f}$ must give $\langle\hat{f}^\dag\hat{f}\rangle\ge0$ for all quantum states. 
Furthermore, if the state remains nonnegative under partial transposition (PT), we also require $\langle\hat{f}^\dag\hat{f}\rangle_{\rm PT}\ge0$ \cite{Vogel}. 
Below, we use the general relation 
\begin{eqnarray}
 \left\langle \hat{O}_A\hat{O}_B \right\rangle_{\rho^{PT}}=\left\langle \hat{O}_A\hat{O}_B^{\dag*} \right\rangle_{\rho},
 \label{eqn:PT-rule}
\end{eqnarray}
where $\hat{O}_A$ and $\hat{O}_B$ are operators acting on subsystems $A$ and $B$, respectively, with PT taken for $B$ \cite{Nha2}. 
The symbol $^*$ denotes complex conjugation of matrix elements.

First, taking $\hat{f}=a+b\hat{C}_1\hat{C}_2$, the condition $\langle\hat{f}^\dag\hat{f}\rangle\ge0$ must be satisfied for arbitrary $a$ and $b$, 
which gives $\left|\langle \hat{C}_1\hat{C}_2\rangle\right|^2\le\langle \hat{C}_1^\dag\hat{C}_1\hat{C}_2^\dag\hat{C}_2\rangle$ for all quantum states---therefore, no violation at all. 
In contrast, the PT condition $\langle\hat{f}^\dag\hat{f}\rangle_{\rm PT}\ge0$ with $\hat{f}=a+b\hat{C}_1\hat{C}_2^*$ gives the inequality~(\ref{eqn:1st-Bell-Q}). 
That is, if the Bell inequality~(\ref{eqn:1st-Bell-Q}) is violated, the state must be NPT.
Secondly, with $\hat{f}=a\hat{C}_1+b\hat{C}_2$, the condition $\langle\hat{f}^\dag\hat{f}\rangle\ge0$ gives $\left|\langle \hat{C}_1\hat{C}_2^\dag\rangle\right|^2\le\langle \hat{C}_1^\dag\hat{C}_1\rangle\langle \hat{C}_2^\dag\hat{C}_2\rangle$ for all quantum states. 
In contrast, its PT version with $\hat{f}=a\hat{C}_1+b\hat{C}_2^*$ gives the inequality~(\ref{eqn:2nd-Bell-Q}), so its violation again confirms NPT entanglement.
The Peres conjecture \cite{Peres, Acin, Nha2} that only NPT entanglement is incompatible with LHV descriptions is thus supported in our framework.

{\it CV case}---Let us first apply the inequalities~(\ref{eqn:1st-Bell-Q}) and~(\ref{eqn:2nd-Bell-Q}) to two-mode CV states. 
The simplest among all possible tests is to take $\hat{C}_j=\hat{a}_j$ ($j=1,2$), where $\hat{a}_j$ is the annihilation operator describing the field amplitude of mode $j$. 
It can be decomposed into two Hermitian operators, $\hat{a}_j=\hat{X}_j+i\hat{Y}_j$ ($j=1,2$), where $\hat{X}_j\equiv\frac{1}{2}(\hat{a}_j+\hat{a}_j^\dag)$ and 
$\hat{Y}_j\equiv\frac{1}{2i}(\hat{a}_j-\hat{a}_j^\dag)$ are two orthogonal quadrature amplitudes. 
Thus, to test the 1st-inequality~(\ref{eqn:1st-Bell-Q}), which reads
\begin{eqnarray}
|\langle \hat{a}_1\hat{a}_2^\dag\rangle|^2 \le \langle \hat{N}_1\hat{N}_2\rangle\hspace{0.3cm}:{\rm 1st-inequality},
\label{eqn:1st-Bell-QO}
\end{eqnarray} 
with $\hat{N}_j=\hat{a}_j^\dag\hat{a}_j$ ($j=1,2$),
the field amplitude correlation $|\langle \hat{a}_1\hat{a}_2^\dag\rangle|^2$ can be measured in 4 segments importantly by {\it local} homodyne measurements, 
\begin{eqnarray}
|\langle \hat{a}_1\hat{a}_2^\dag\rangle|^2=\left(\langle \hat{X}_1\hat{X}_2\rangle+\langle \hat{Y}_1\hat{Y}_2\rangle\right)^2+\left(\langle \hat{X}_1\hat{Y}_2\rangle-\langle \hat{Y}_1\hat{X}_2\rangle\right)^2.\nonumber\\
\end{eqnarray}
On the other hand, the intensity correlation $\langle \hat{N}_1\hat{N}_2\rangle$ can be measured by photon counting at each mode.
There are some broad classes of two-mode states that violate the inequality~(\ref{eqn:1st-Bell-QO}). 
The most practically feasible among them is the single-photon entangled state, $|\Psi_s\rangle=\cos\theta|1,0\rangle+\sin\theta e^{-i\phi}|0,1\rangle$, 
giving $|\langle \hat{a}_1\hat{a}_2^\dag\rangle|^2=\frac{1}{4}\sin^22\theta$ and $\langle \hat{N}_1\hat{N}_2\rangle=0$.
In view of~(\ref{eqn:1st-Bell-QO}), the degree of violation can be measured by the deviation of the ratio $\frac{\rm LHS}{\rm RHS}=\frac{|\langle \hat{a}_1\hat{a}_2^\dag\rangle|^2}{\langle \hat{N}_1\hat{N}_2\rangle}$ from unity, which becomes infinite in this case.

We note that the same inequality~(\ref{eqn:1st-Bell-QO}) was derived also by Hillery and Zubairy \cite{Hillery}, but within a distinct context of
separability condition along a different route for non-Gaussian entanglement \cite{Nha3,Agarwal}. 
The proposed schemes to test the inequality~(\ref{eqn:1st-Bell-QO}) in \cite{Hillery, Nha3} considered collective measurements from the SU(2) algebra, not local ones as proposed here, thus unsuitable for nonlocality test. 
%Nevertheless, other physical systems (e.g. quantum-beat laser \cite{Zubairy}) known to fulfill the HZ criterion may also provide realizable experimental tests of CV nonlocality.   
A closely-related Bell inequality was also derived by Cavalcanti {\it et al.} in \cite{Cavalcanti}, where the intensity correlation $\langle |C_1|^2|C_2|^2\rangle$ in~(\ref{eqn:Bell2}) was addressed by the squared-quadrature correlation, $\langle (\hat{X}_1^2+\hat{Y}_1^2)(\hat{X}_2^2+\hat{Y}_2^2)\rangle$. 
Thus, they considered only the wave-like correlations in both sides \cite{note0}, and it is known that no violation occurs for two-mode states within their framework \cite{Acin}. In contrast, our inequality incorporates two distinct aspects of field correlations in~(\ref{eqn:1st-Bell-QO}), the wave-like (LHS) and the particle-like (RHS) correlation. 
In this sense, our approach emphasizes the role of the wave-particle dual aspects in manifesting the failure of local realism \cite{note}.

Let us turn our attention to the 2nd-inequality~(\ref{eqn:2nd-Bell-Q}). 
Again, by the substitution $\hat{C}_j=\hat{a}_j$ ($j=1,2$), we obtain
\begin{eqnarray}
|\langle \hat{a}_1\hat{a}_2\rangle|^2 \le \langle \hat{N}_1\rangle\langle \hat{N}_2\rangle\hspace{0.3cm}:{\rm 2nd-inequality},
\label{eqn:2nd-Bell-QO}
\end{eqnarray} 
which also appeared as a separability condition in \cite{Hillery}. 
In this case, the optical EPR state (two-mode squeezed state), $|{\rm TMSS}\rangle=e^{r(a_1^\dag a_2^\dag-a_1a_2)}|0,0\rangle=\sum_{n=0}^\infty\frac{\tanh^nr}{\cosh r}|n,n\rangle$, violates the inequality regardless of $r$  (degree of squeezing). 
The degree of violation measured by $\frac{|\langle a_1a_2\rangle|^2}{\langle a_1^\dag a_1\rangle\langle a_2^\dag a_2\rangle}-1=\tanh^{-2}r-1$ ($r>0$) 
increases with $r$ decreasing and becomes extremely large as $r\rightarrow 0$. 
%(Of course, no violation occurs at $r=0$). 

{\it Loophole-free test}---Let us now address how our tests can avoid both the locality and the detector-efficiency loopholes. 
First, to enforce a strict locality condition, a random-number generator yielding $R=0,1,$ and 2 can be used at each observer to choose local measurement settings, 
similar to the method of \cite{Weihs}. In the balanced homodyne detection, the local oscillator (LO) with adjustable phase is mixed with a signal at a 50:50 beam-splitter (BS). In our case, an electronic attenuator can be put between the LO and the BS to reduce/unblock the LO. For $R=0$ ($R=1$) case, the LO phase is adjusted to $X$ ($Y$) quadrature with the attenuator off (homodyne detection). For $R=2$, the attenuator turns on to reduce the LO, measuring the signal intensity (photon counting) as below. 
If $R$ is randomly generated at the last instant when the signal impinges on the BS, the time-like communication between two observers can be ruled out \cite{Weihs}. 

Second, we consider a full LHV model including all non-detection events to address the detection-loophole issue. For the case that the real or the imaginary part of  $C_j\equiv C_{jx}+iC_{jy}$ ($j=1,2$) is undetected, one may assign a fixed value 0 to such events and the inequality~(\ref{eqn:Bell1}) still holds.
The intensities of the RHS can be decomposed as $\langle|C_j|^2\rangle=p_{j,D}\langle|C_j|^2\rangle_D+(1-p_{j,D})\langle|C_j|^2\rangle_U$ for the 2nd-inequality~(\ref{eqn:2nd-Bell-Q}), 
where $\langle|C_j|^2\rangle_{D,U}$ denotes the intensity average for detected/undetected ensembles, with $p_{j,D}$ the detection probability in photon counting. 
A consequence of nonideal efficiency $\eta<1$ is $\langle|C_j|^2\rangle_U\le\langle|C_j|^2\rangle_D$, which can be proved within classical description. In turn, it gives
$\langle|C_j|^2\rangle_U\le\langle|C_j|^2\rangle$, where $\langle|C_j|^2\rangle=\langle C_{jx}^2+C_{jy}^2\rangle$ is the total intensity average that can be alternatively measured via homodyne detection. 
Therefore, a full LHV inequality leads to
$|\langle \hat{a}_1\hat{a}_2\rangle|^2 
\le\prod_{j=1,2}\left[p_{j,D}\langle \hat{N}_j\rangle_D+(1-p_{j,D})\langle \hat{X}_{j}^2+\hat{Y}_{j}^2\rangle\right]\nonumber$.
To enhance the detection probability $p_{j,D}$, one may mix the signal with LO (amplitude$\sim\beta$)  at a beam splitter and measure the intensity sum ${\cal S}$ of two outputs. In each event, the signal intensity is assigned the value ${\cal S}-\langle I\rangle_{LO}$ where $\langle I\rangle_{LO}$ is the LO intensity average that can be separately measured. 
Our LHV inequalities are still valid with the LO field included as another (predetermined) random variable, and  $p_{j,D}$ rapidly approaches 1 by increasing $\beta$ for any $\eta$ and squeezing $r$. In this case, the contribution of the second term $\langle \hat{X}_{j}^2+\hat{Y}_{j}^2\rangle$ is negligible.

The photodetection with efficiency $\eta$ is practically equivalent to the ideal detection after the signal $\hat{a}_j$ is mixed with a vacuum $\hat{v}_j$ at a beam-splitter of transmissivity $\sqrt{\eta}$. Namely, the observed signal $\hat{a}_{oj}$ is expressed by  $\hat{a}_{oj}=\sqrt{\eta}\hat{a}_j+\sqrt{1-\eta}\hat{v}_j$ ($j=1,2$). This gives  $\langle \hat{N}_{o1}\hat{N}_{o2}\rangle=\eta^2\langle \hat{N}_1\hat{N}_2\rangle$ and 
$\langle \hat{N}_{o1}\rangle\langle \hat{N}_{o2}\rangle=\eta^2\langle \hat{N}_1\rangle\langle \hat{N}_2\rangle$ in the above-mentioned intensity measurement. 
In the balanced homodyne detection to measure quadrature amplitudes $\hat{X}_j$ and $\hat{Y}_j$ at each mode, the same model applies, $\hat{a}_{oj}=\sqrt{\eta}\hat{a}_j+\sqrt{1-\eta}\hat{v}_j$, in the limit of large-intensity local oscillator \cite{Leonhardt}. 
This gives $|\langle \hat{a}_{o1}\hat{a}_{o2}\rangle|^2=\eta^2|\langle \hat{a}_1\hat{a}_2\rangle|^2$ and  $|\langle \hat{a}_{o1}\hat{a}_{o2}^\dag \rangle|^2=\eta^2|\langle \hat{a}_1\hat{a}_2^\dag\rangle|^2$. 
Therefore, $\eta^2$ becomes an overall factor in both sides of~(\ref{eqn:1st-Bell-QO}) and~(\ref{eqn:2nd-Bell-QO}), which makes our scheme insensitive to detector efficiency. 
%If the efficiency is low, it would of course take longer time to obtain substantial data, however, it would not affect the degree of violation
%measured by $\frac{\rm LHS}{\rm RHS}-1$.

%An experimental error may also occur in spatially distributing a two-mode state to each observer. When each mode is exposed to a vacuum-reservoir, it undergoes dissipation and decoherence. The vacuum-reservoir can be modeled precisely by a beam-splitter setting \cite{Kim}, thus the above analysis also applies here, implying that the environmental effects again cancel in both sides of~(\ref{eqn:1st-Bell-QO}) and~(\ref{eqn:2nd-Bell-QO}).  
%In particular, the single-photon entangled state, $|\Psi_s\rangle$, can be generated by mixing a single-photon with a vacuum at a beam-splitter of transmissivity $\cos\theta$. In realistic conditions, the single-photon is produced in a mixed state,  
%$\rho_{\rm single}=p_S|1\rangle\langle1|+(1-p_S)|0\rangle\langle0|$, with the success probability $p_S$ \cite{Lvovsky}. 
%The output two-mode state from the beam splitter becomes $\rho=p_S|\Psi\rangle\langle \Psi|+(1-p_S)|0,0\rangle\langle0,0|$, 
%and the degree of violation in~(\ref{eqn:1st-Bell-QO}) is still very large with $\langle \hat{N}_1\hat{N}_2\rangle=0$ for any $p_S>0$.

{\it Multipartite systems}---The inequalities~(\ref{eqn:1st-Bell-Q}) and~(\ref{eqn:2nd-Bell-Q}) can be further generalized to $N$-partite systems as 
\begin{eqnarray}
\left|\left\langle \prod_{i=1}^k\hat{C}_i\prod_{j=k+1}^N\hat{C}_j^\dag\right\rangle\right|^2\le\left\langle \prod_{i=1}^N\hat{C}_i^\dag\hat{C}_i\right\rangle\hspace{0.1cm}:{\rm 1st},
\label{eqn:1st-Bell-MQ}
\end{eqnarray}
\begin{eqnarray}
\left|\left\langle \prod_{i=1}^N\hat{C}_i\right\rangle\right|^2\le\left\langle \prod_{i=1}^k\hat{C}_i^\dag\hat{C}_i\right\rangle\left\langle \prod_{j=k+1}^N\hat{C}_j^\dag\hat{C}_j\right\rangle\hspace{0.05cm}:{\rm 2nd},\nonumber\\
\label{eqn:2nd-Bell-MQ}
\end{eqnarray}
where $N$-parties are divided into two groups of $k$- and $N-k$ modes $(k=1,\ldots,N-1)$.  
It is again readily proved that only NPT entangled states can violate the inequalities~(\ref{eqn:1st-Bell-MQ}) and~(\ref{eqn:2nd-Bell-MQ}): Take $\hat{f}=A+B\prod_{i=1}^k\hat{C}_i\prod_{j=k+1}^N\hat{C}_j^*$ and $\hat{f}=A\prod_{i=1}^k\hat{C}_i+B\prod_{j=k+1}^N\hat{C}_j^*$, respectively, for the condition $\langle\hat{f}^\dag\hat{f}\rangle_{\rm PT}\ge0$.
%Although the violation of~(\ref{eqn:1st-Bell-MQ}) and~(\ref{eqn:2nd-Bell-MQ}) can also be demonstrated by higher-order generalizations, $\hat{C}_j=\hat{a}_j^{m_j}$ ($m_j$: positive integer), let us here consider the simplest case $m_j=1$ for all modes. 
The $N$-mode GHZ state with a mixture of vacuum, $\rho=p_S|{\rm GHZ}\rangle\langle {\rm GHZ}|+(1-p_S)|0\cdots0\rangle\langle0\cdots0|$,  
where $|{\rm GHZ}\rangle=c_1|\{1\}_k,\{0\}_{N-k}\rangle+c_2|\{0\}_k,\{1\}_{N-k}\rangle$ is the superposition of  $k$ modes ($N-k$ modes) all occupying one (no) photon and vice versa, violates the inequality~(\ref{eqn:1st-Bell-MQ}). 
The multimode EPR state, produced by injecting single-mode squeezed states into a series of beam splitters \cite{Braunstein}, violates the inequality~(\ref{eqn:2nd-Bell-MQ}). 
The violation can occur with any partition numbers $(k=1,\ldots,N-1)$, thereby manifesting the true multipartite nature of nonlocality to some extent. %Further details will be given elsewhere.
%$B=\prod_{j=1}^{N-1}B_{j+1j}\left(\cos^{-1}\frac{1}{\sqrt{N-j+1}}\right)$, where $B_{ij}(\theta)$ is the beam-splitter action between two modes $i$ and $j$ with the transmissivity  $\cos\theta$ \cite{Braunstein}.
%The most feasible multipartite EPR-type state can be generated by injecting only 1-mode squeezed state and $n-1$-mode vacuum states to the beam-splitter array \cite{Braunstein}. The $n$-mode EPR state then violates the inequality~(\ref{eqn:2nd-Bell-MQ}) for any bipartitions with even number $n$. %For example, the 4-mode state yields the ratio $\frac{\rm LHS}{\rm RHS}=\frac{3}{\tanh^2r(3+2\tanh^2r)}$ and $\left(\frac{3}{1+2\tanh^2r}\right)^2$ for the partition $k=1,2$, respectively, thus violating the Bell inequality for any bipartitions with input squeezing $r<r_c\approx1.183$.
%\frac{|\left\langle a_1\cdots a_4\right\rangle|^2}{\left\langle a_1^\dag a_1a_2^\dag a_2\right\rangle\left\langle a_3^\dag a_3a_4^\dag a_4\right\rangle}
%In the multimode tests, the robustness against experimental imperfections emerge as before. 
%In the higher-order test $m_i>1$ for some modes$i$, however, it is not clear how one can generally come up with {\it local} measurements to detect the field correlation $
%\left|\left\langle \prod_{i=1}^k\hat{C}_i\prod_{j=k+1}^N\hat{C}_j^\dag\right\rangle\right|^2$, 
%although there exists a proposal to indirectly measure such multi-mode correlations \cite{Vogel1}.

In summary, we derived two classes of multi-mode Bell inequalities that can be greatly violated by readily accessible $n$-mode CV states for $n\ge2$. 
Remarkably, our proposed tests are insensitive to detector-efficiency making a loophole-free test very feasible within existing technologies. 
The role of wave and particle correlations is highlighted particularly for single-photon nonlocality \cite{Tan}, and this may suggest possibilities for addressing Bell theorem and related issues in a new perspective.

The authors thank M.S. Kim for useful comments, and J. Lee, K. Modi, and Ed Fry for discussions. 
This work is supported by NPRP 1-7-7-6 from QNRF and IT R\& D program of MKE/KEIT (KI001789). 
\newline \newline 
*corresponding author: hyunchul.nha@qatar.tamu.edu


\begin{thebibliography}{99}
\bibitem{EPR} A. Einstein, B. Podolsky, and N. Rosen, Phys. Rev. {\bf 47}, 777 (1935). 
\bibitem{Bell} J.S. Bell, Physics (Long Island City, N.Y.) {\bf 1}, 195 (1964); J. F. Clauser {\it et al.}, \prl {\bf 23}, 880 (1969).
\bibitem{Ekert} A. K. Ekert, \prl {\bf 67}, 661 (1991); A. Acin, N. Gisin, and L. Masanes, {\it ibid} {\bf 97}, 120405 (2006).
\bibitem{Weihs} G. Weihs {\it et al.}, \prl {\bf 81}, 5039 (1998); M. A. Rowe {\it et al.}, Nature {\bf 409}, 791 (2001); D. N. Matsukevich {\it et al.}, \prl {\bf 100}, 150404 (2008).
\bibitem{Fry} E. S. Fry, T. Walther, and S. Li,  \pra {\bf 52}, 4381 (1995); A. Cabello, D. Rodriguez, and I. Villanueva, \prl {\bf 101}, 120402 (2008).
\bibitem{Ansmann} M. Ansmann {\it et al.}, Nature {\bf 461}, 504 (2009); E. S. Fry and R. C. Thompson, \prl {\bf 37}, 465 (1976); 
A. Aspect, P. Grangier, and G. Roger, {\it ibid} {\bf 47}, 460 (1981).
\bibitem{Bell1}J. S. Bell, {\it Speakable and Unspeakable in Quantum Mechanics} (Cambridge University Press, 1987). 
\bibitem{Banaszek} K. Banaszek and K. Wodkiewicz, \pra {\bf 58}, 4345 (1998); \prl {\bf 82}, 2009 (1999). 
\bibitem{Chen} Z.-B. Chen {\it et al.}, \prl {\bf 88}, 040406 (2002)
\bibitem{Nha1} H. Nha and H. J. Carmichael, \prl {\bf 93}, 020401 (2004);  R. Garcia-Patron {\it et al.}, \prl {\bf 93}, 130409 (2004).
\bibitem{Acin0} A. Acin {\it et al.}, \pra {\bf 79}, 012112 (2009).
\bibitem{Vogel} E. Shchukin and W. Vogel, \prl {\bf 95}, 230502 (2005). 
It provides a CV framework that distinguishes separable and entangled states {\it within} quantum formulation. 
In contrast, Bell inequalities are to distinguish classical LHV and other (quantum) descriptions of correlation.
\bibitem{Cavalcanti} E. Cavalcanti {\it et al.}, \prl {\bf 99}, 210405 (2007). %C. J. Foster, M. D. Reid, and P. D. Drummond
\bibitem{He} Q. Y. He {\it et al.}, \prl {\bf 103}, 180402 (2009).
%\bibitem{He} Q. Y. He, E. G. Cavalcanti, M. D. Reid, and P. D. Drummond, \prl {\bf 103}, 180402 (2009).
\bibitem{Acin} A. Salles, D. Cavalcanti, and A. Acin, \prl {\bf 101}, 040404 (2008); A. Salles {\it et al.}, Quantum. Inf. Comput. {\bf 10}, 0703 (2010).
\bibitem{Peres} A. Peres, Found. Phys. {\bf 29}, 589 (1999).
\bibitem{Nha2} Q. Sun, H. Nha, and M. S. Zubairy, \pra {\bf 80}, 020101(R) (2009).
\bibitem{Hillery} M. Hillery and M. S. Zubairy, \prl {\bf 96}, 050503 (2006); \pra {\bf 74}, 032333 (2006); M. Hillery, H. T. Dung, and J. Niset, \pra {\bf 80}, 052335 (2009).
\bibitem{Nha3} H. Nha and J. Kim, \pra {\bf 74}, 012317 (2006). 
\bibitem{Agarwal} G. S. Agarwal and A. Biswas, New J. Phys. {\bf 7}, 211 (2005). 
%\bibitem{Zubairy} S. Qamar {\it et al.}, \pra {\bf 77}, 062308 (2008).
\bibitem{note0} The identity $\hat{X}_j^2+\hat{Y}_j^2=\hat{a}_j^\dag\hat{a}_j+\frac{1}{2}$ may suggest the measurement of intensity correlation for the inequality in \cite{Cavalcanti} also by photon counting. However, note that the numeric constant $\frac{1}{2}$ here, originating from the quantum commutator, does not reflect realistic photon-counting events.
\bibitem{note} As the violation of~(\ref{eqn:1st-Bell-QO}) occurs with three local measurements (quadratures ${\hat X}_j,\hat{Y}_j$, and intensity $\hat{Z}_j=a_j^\dag a_j$), the LHV description might accordingly invoke three random variables $x_j,y_j$ and $z_j\ge0$ to explain the violation. However, there exists a natural classical constraint, $z_j=x_j^2+y_j^2$ (intensity), which again leads to the inequality~(\ref{eqn:Bell2}).  
The same logic also applies to~(\ref{eqn:2nd-Bell-QO}). 
\bibitem{Leonhardt} U. Leonhardt and H. Paul, \pra {\bf 48}, 4598 (1993). 
%\bibitem{Kim} M. S. Kim and N. Imoto, \pra {\bf 52}, 2401 (1995).
%\bibitem{Lvovsky} A.~I.~Lvovsky {\it et al.}, \prl {\bf 87}, 050402 (2001); 
%A.~I.~Lvovsky and J.~Mlynek, {\it ibid.} {\bf 88}, 250401 (2002).
\bibitem{Braunstein} P. van Loock and S. Braunstein, \prl {\bf 84}, 3482 (2000). 
%\bibitem{Vogel1} E. Shchukin and W. Vogel, \prl {\bf 96}, 200403 (2006). 
\bibitem{Tan} S. M. Tan, D. F. Walls, and M. J. Collett, \prl {\bf 66}, 252 (1991); L. Hardy, {\it ibid.} {\bf 73}, 2279 (1994); 
B. Hessmo {\it et al.}, {\it ibid.} {\bf 92}, 180401 (2004); 
J. Dunningham and V. Vedral, {\it ibid.} {\bf 99}, 180404 (2007).
\end{thebibliography}
\end{document}